\documentclass[a4paper,pra,10pt,byrevtex,titlepage,secnumarabic,twocolumn,nofootinbib,longbibliography,showpacs,showkeys]{revtex4-1}
\usepackage{graphicx}
\usepackage{draftcopy}
\usepackage{amsmath}
\usepackage{amssymb}
\usepackage{booktabs}
\usepackage{url}
\usepackage[squaren]{SIunits}
\usepackage[utf8]{inputenc}
\usepackage{array}
\newcommand{\uhora}[1]{\mbox{#1}}
\begin{document}
\author{José María Martín Olalla}
\email{olalla@us.es}
\homepage{Twitter: @MartinOlalla\_JM}
\affiliation{Departamento de Física de la Materia Condensada. Universidad de Sevilla. Ap Correos 1065, ES41080. Sevilla. Spain}
\title{The winter day as an constraint for human activity in Western Europe}

\published{\today}

\begin{abstract}
Time use surveys in Denmark, Spain, France, Ireland, Italy and United Kingdom are analyzed to provide start, noon and end times for the main activities of a society: labor (the focus of this preprint), sleeping and eating. Also, the location at home is analyzed. Local times are converted into mean solar times and compared to latitude. Observed trends allow to unveil the winter day as a restriction for the human activity. Alternatively, apparently large time differences set forth by clocks, becomes smaller when observed as a time distance to winter sunrise or sunset. 
\end{abstract}
\maketitle

\section{Introduction}
\label{sec:intro}
Time use surveys characterize the behavior of societies by computing the amount of time individuals spent in labor, household activities, leisure activities and sleeping on an average day. They also provide valuable information on how these activities are shared by gender or age. 

Most of the information disseminated by Time use surveys concerns intervals of time expressed as an average number of hours per day. Yet this work focuses on the times at which main activities (eating, sleeping and working) occur. Time use surveys also locate where an activity is being held from which it is possible to track the binary condition of being at home or out of home.

This work is a follow up of a previous preprint\cite{2014arXiv1406.4763M} focused on Time Use Surveys in Italy\cite{ittus-2010}, Spain\cite{estus-2010} and United Kingdom\cite{uktus-2003}. Here surveys from Denmark\cite{dktus-2001}, France\cite{frtus-2010} and Ireland\cite{ietus-2005} will also be analyzed. The methodology (see Section~\ref{sec:methodology}) slightly differs from the previous work which helps improving the analysis of the afternoon. As in the previous work the microdata of the surveys here analyzed were retrieved freely from the Internet (Spain and Denmark) or were obtained after contacting with the institution. That was the single criterion to pick a survey.

Irish National Time Use Survey slices the day in ninety six slots of fifteen minutes in duration. Every other survey slices the day in one hundred and forty four slots of ten minutes in duration. The respondent is compelled to fill the relevant information in every spot. That is: which activity he was doing, where, and with whom, among others. The questionnaires also collect a myriad of information as for instance week day and month of the diary, gender of the respondent, incomes or number of individuals in the household. 

The burden of this paper is linking the social activity to the solar activity in Central and Western Europe.  For so doing we will compare solar times at which activities occur rather than local times which are computed from the surveys. 

It will be shown that the winter day, which yields the latest sunrise and the earliest sunset of the year, is a powerful restriction for the social activities and provide an explanation for the way in which they occur all the year round. Notice that winter day length depends solely on the latitude.

\section{Methodology}
\label{sec:methodology}

\subsection{Definitions}
\label{sec:definitions}

Only respondents who reported labor activity at some point of the day in a week day (Monday to Friday) will be analyzed.

Daily reports will be grouped along the First Level of the Nomenclature of Territorial Units for Statistics (NUTS-1) which is described elsewhere.\cite{nuts-1-wiki} When  diaries are geolocated to the NUTS-2 level, data will be weight-averaged by population to get NUTS-1 level statistics. Finally, the weight-averaged value of NUTS-1 level statistics will be the country level mean result.

This work analyzes daily activity plot which, essentially, computes the shares of people doing an activity as a function of time and during an average week day. This work will not thoroughly describe labor activities and their differences from region to region or from country to country, yet a brief description of the labor activity is worth to quote.

Labor activity soars after the minimum level in the morning. Then, somewhere close to noon the activity attains a maximum and starts decreasing, that is lunch time. Then labor activity is resumed, and gets a relative maximum in the afternoon. Then it starts decreasing to the ground level. Notice that the area enclosed by the labor activity plot has units of time and indeed is the mean daily amount of time spent at work by an employed person.

Every labor activity plot displays this simple behavior with  forenoon and afternoon peaks, sharp up-rise in the morning and a slower decay in the afternoon. Not surprisingly, magnitudes like the strength of the peaks or their relative position regionally changes.

Labor activity plot will then be characterized by three times: start time, noon time and end time. Noon time is easily computed as the time when half the total area has been consumed. The inset of Fig.~\ref{fig:labor} display the position of these times for the labor activity displayed.

Start and end times essentially requires the definition of a threshold level above which one may reasonable say that the region is ``working'' and below which one may reasonable say that the region is not ``working''. Notice that this approach slightly differs from statistically analyzing the distribution of times when individuals get to work or leave working.\cite{2014arXiv1406.4763M}

A very simple choice for a threshold is the 50\% level, but this choice jeopardizes the analysis in some regions (notably in Spain and Southern Italy) because labor activity is not resumed back to the 50\% in the afternoon. Also, it is never the case that 100\% of the employed persons are on duty. Labor activity only reaches the 90\% level in some regions of Italy. In contrast some regions only reaches the 70\% level.

A naive election for the threshold is then half the value of strongest peak. Which is to say the 50\% of the peak. Notice that this threshold, though variable, is consistently set for every region. Also notice that this threshold is always weaker than the choice of the 50\% level, thus leading to earlier start times and later end times.

The threshold allows us to introduce the labor start time ---computed as the time when labor activity overshoots the threshold in the morning--- and the labor end ---computed as the time when labor activity undershoots the threshold in the afternoon.

Likewise, activity plots can be retrieved for the location at-home (see Appendix~\ref{sec:at-home}), the eating activity (see Appendix~\ref{sec:eating-times}) and the sleeping activity (see Appendix~\ref{sec:wakeful-times}). For the sleeping and at-home analysis  the threshold level is effectively set to the 50\% since peaks gets close to the 100\% level: almost every employee is awaken or out of home at some specific time.

On the other hand, the eating activity is different since it does not smoothly change on time. Instead, it occurs in bursts: in the morning ---breakfast---, noon ---lunch--- and afternoon ---dinner. Eating times will be ascribed to the time position of the peak for every burst.

\subsection{Comparing times}
\label{sec:comparing-times}

The burden of this work is to provide a comprehensive comparison of the times when the main activities occur.

Results obtained from Time Use Survey only display local times. This is a very awful quantity to compare because it lacks a common time basis. To be more specific: they depend on the local time zone. 

Reducing data to a common time basis (say, for instance, UTC) will only display Earth's rotation with eastern regions displaying earlier times than western regions.

The only comprehensive way of comparing social time over a wide geographical areas is through the local mean solar time which accounts for the local time zone $\Delta_i$ and Earth's rotation. Local mean solar time $t^{\star}_i$ given by:
\begin{equation}
  \label{eq:1}
  t^{\star}_i=t_i - \Delta_i + \lambda_i \omega^{-1}
\end{equation}
where $t_i$ is any time defined in Sec.~\ref{sec:definitions} expressed in local time, $\Delta_i$ is the local time zone expressed as an offset to UTC time, $\lambda_i$ is the mean longitude of the region, expressed as an offset to the UTC prime meridian (Greenwich meridian) and $\omega$ is Earth's angular rotation speed (one revolution per day, conveniently expressed as $\unit{15}{\degree\usk\reciprocal\hour}$ in this context). The quantity
\begin{equation}
  \label{eq:2}
\tau_i=\Delta_i-\lambda_i\omega^{-1},
\end{equation}
 is the time offset between local civil time and local solar time. It generally ranges in $\pm\unit{30}{\minute}$. This range is exceeded when local time zone is not the geographical time zone, as in the case of Spain and France since end of World War Two.

Mean solar local times $t^{\star}_i$ can be directly linked to solar activity. The most simple case is solar noon which, by definition, occur at $\tau=\unit{12}{\hour}$ everywhere but it is also the case for properties like sunrises and sunsets.

\subsection{Geolocation}
\label{sec:geolocation}

As it was mentioned earlier on Sec.~\ref{sec:methodology} statistic data from Time Use Survey will be grouped along NUTS-1 regions. In Eq.~(\ref{eq:1}) the longitude of the region is needed. In the forthcoming discussion its latitude will also be employed. This geographical data will be retrieved from database after weight-averaging geographical coordinates of cities whose population is large than 1000 people and exists on the database. This will suffice for the purpose of this paper.

\begin{figure}
  \centering
  \includegraphics[width=\linewidth]{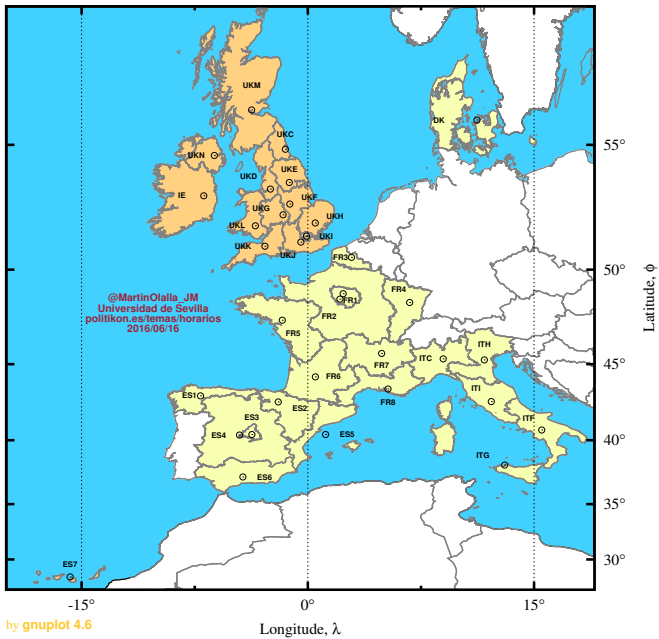}
  \caption{A map of Central and Western Europe. Shaded NUTS-1 regions are analyzed in this work. Darker regions (Ireland, United Kingdom and Canary Islands (ES7)) set clocks to Western European Time zone (WET=UTC+00), lighter regions set clocks to Central European Time zone (CET=UTC+01).}
  \label{fig:mapaeuropa}
\end{figure}

Fig.~\ref{fig:mapaeuropa} shows the mean values of the geographical data obtained from the database and the weighted-average values (open symbols) corresponding to the country level longitude and latitude. Notice that for computing this country level mean values ES7 is not considered in Spain, and overseas departments are not considered in France.

\section{Results}
\label{sec:results}

\begin{figure*}
  \centering
  \includegraphics[width=\textwidth]{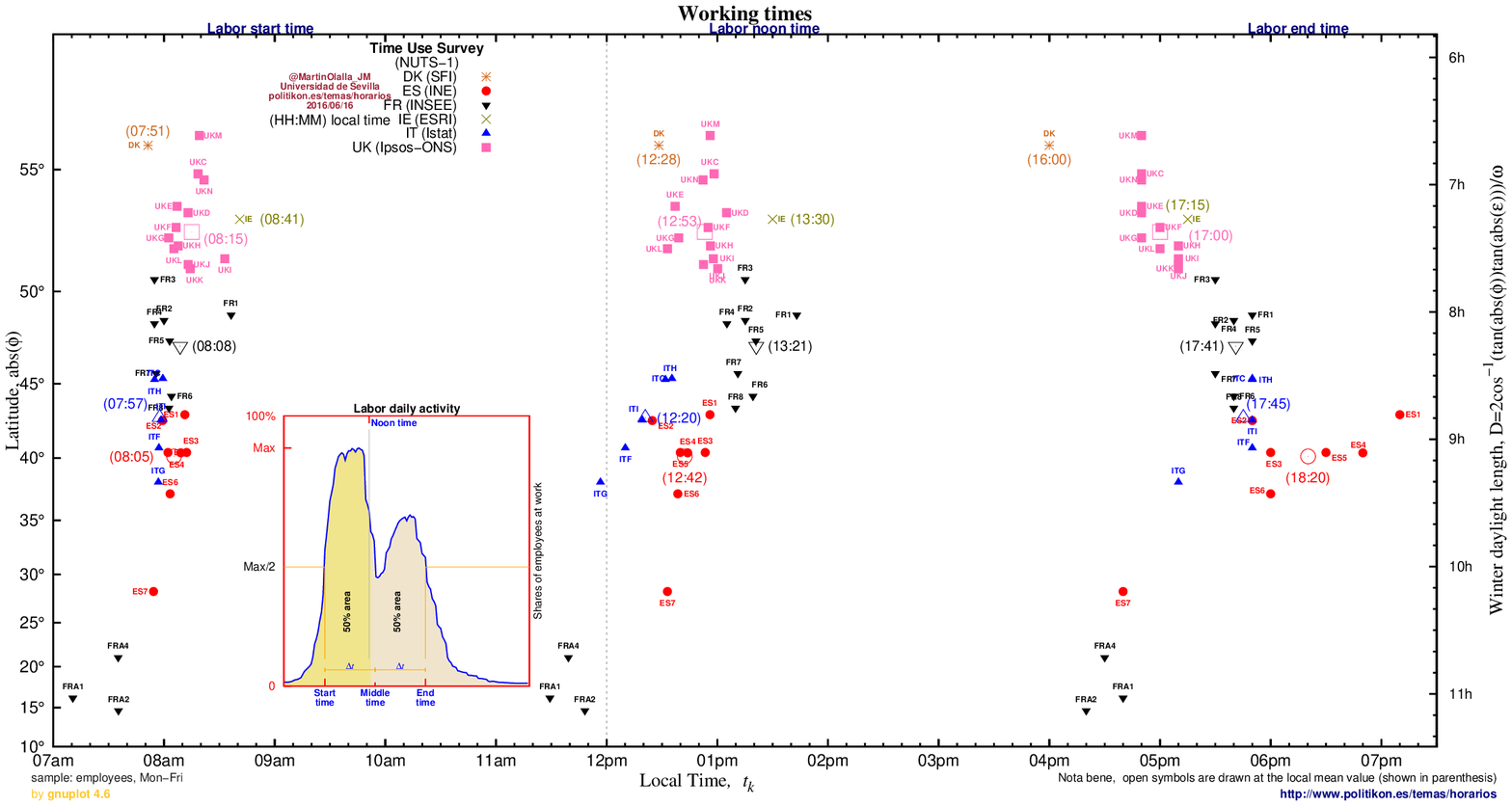}

  \includegraphics[width=\textwidth]{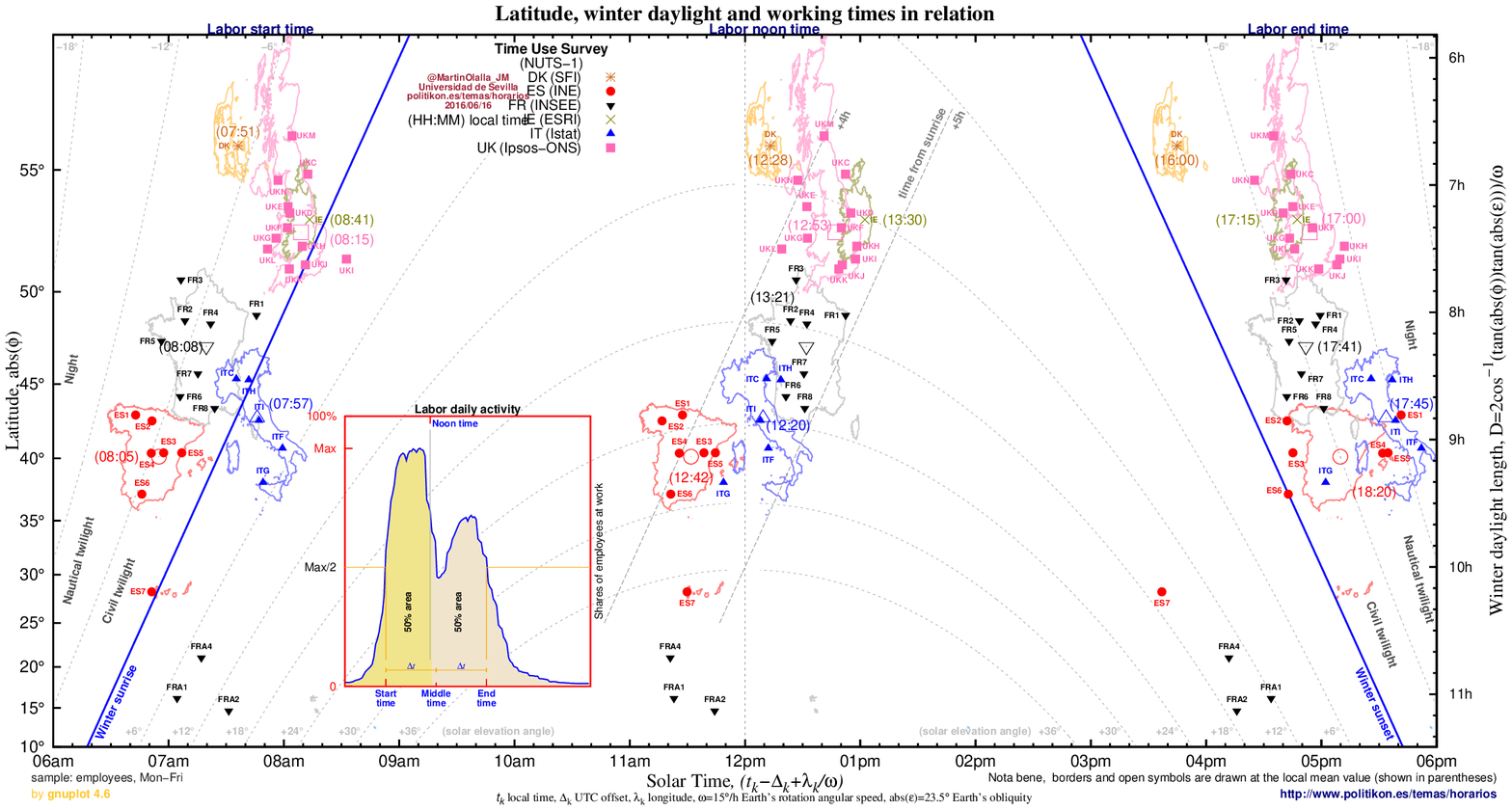}
  \caption{Top figure: labor start, noon and end local times plotted versus latitude. Morning times looks like similar across surveys with \uhora{08}am as a popular choice in Europe. Labor end times shows a trend with later end times as latitude decreases. The inset shows an labor activity plot and the definition of the labor start, noon, and end times. Labor middle times (not shown in the figure) are mostly independent of latitude. Bottom figure: the same as before but with the $x$-axis displaying solar times instead of local times. Solar activity and borders can be added to the plot, including winter sunrise and sunset lines. Labor times correlated with the winter day.}
  \label{fig:labor}
\end{figure*}

The top panel of Fig.~\ref{fig:labor} shows the latitude in the vertical axis versus start, noon and end labor local times for the set of regions analyzed in this work.

In the morning most of data are grouped around \uhora{08:00am} local time. In the afternoon labor end times display a downward trend with end times increasing with decreasing latitude. At noon data points seem rather scattered.

Similar results can be observed for the eating activity and the at-home location. As shown in the appendixes.

In the search of an explanation one may think that social activity arises from a fundamental rest state at midnight when people mostly slept\footnote{Albeit for a non-zero background level corresponding to a ``social watch''}. Hence, wake-up time and, more generally, the start times arise from a common initial condition. Hereafter, cultural and legal factors may influence the behavior of the activities and would trigger differences at the end of the day. 

Even if that were the case, two fundamental questions would arise. The first one is  why is eight o'clock the preferred value of the labor start time. The second one is why labor end times follow this specific trend characterized by a unique slope.

A better insight to the problem and an effective answer to these questions is obtained when the local mean solar time (see Eq.~(\ref{eq:1})) is plotted instead of the local time as observed in the bottom panel of Fig.~\ref{fig:labor}. The reader will observe in this plot geopolitical borders, which are plotted at the country-level mean value.\footnote{After setting $t_k$ and since $\Delta_k$ is also set, horizontal axis display longitude while vertical axis display latitude; hence borders can be displayed for this specific $t_k$.}

Upon this choice of $x$-axis, solar properties can be plotted as well. Fig.~\ref{fig:labor} (bottom) shows lines of constant solar elevation angles $z$ including the null condition $z=\unit{0}{\degree}$ which approximately sets the sunrise and sunset lines.\cite{pierre-cam-2014} 

The reader should notice that the winter day is an extremal condition for the solar activity since the sun is apparently traveling around the tropic which is opposite to the observer, yielding to the latest sunrise, the earliest sunset and, also, the smallest solar elevation angle at any time of the day. As a result in the trapeze bounded by parallels (top and bottom) and the sunrise and sunset lines (sides) the solar elevation angle is always positive irrespective of the calendar day. Say, there is always daylight in this area and human activity can be developed at ease.

It should be stressed however that data points do not display winter properties of the distribution of working times but yearly averaged properties.

This analysis provides a simple explanation for the main properties of the labor activity. Labor start time ---year round--- is chosen close to the winter sunrise line. That way, employed persons never get to work too dark at dawn. Labor end times ---year round--- is mostly coincidental to the winter sunset line. Differences may well appear here because of lack of common legal regulations or differences in the shares of economic sectors through survey. Nonetheless, evidence suggests that the winter sunset is also an attractor for the end of the labor activity. That way, as the year progresses, employed persons will mostly enjoy later sunsets in their leisure time.

\section{Discussion}
\label{sec:discussion}

\subsection{Sunrise, sunset, time zone}
\label{sec:geogr-disc}

In understanding this analysis first notice that, in the morning, labor start solar times have a trend with latitude: the more latitude, the later the labor start solar time. In the afternoon the trend is just the opposite. The significant point here is the fact that yearly averaged labor times do not follow yearly averaged solar activity, which is independent of latitude.\footnote{For instance yearly averaged solar sunrise and sunset times are \uhora{06}am and \uhora{06}pm, irrespective of the latitude ---if daylight saving time is not considered. Thus yearly averaged sunrise and sunset local times only depends on longitude for a given time zone.} Modern societies do not filter out the solar seasonal cycle, they have memory.

The trends are easily explained by the solar activity at the winter day which sets the latest sunrise of the year, the earliest sunset of the year and the shortest daylight length. In fact, Fig.~\ref{fig:labor} (top) panel just shows up an interesting natural phenomenon: the way in which the terminator line\footnote{The line separating day from night.} sweeps Europe in winter. This can be easily understood by watching a animation of the winter day in Europe\footnote{\url{http://bit.ly/1U8enMa}} where it can be seen how winter sunrise line sweeps Europe from Southeast to Northwest (the earliest sunrise coming at Southeastern Italy, the latest sunrise at the Scottish Highlands) and it is mostly coincidental over Northwestern Italy, Germany, the French hexagon, Denmark, Netherlands, Belgium and the Iberian Peninsula.\footnote{Winter sunrise in Berlin and Barcelona are coincidental. Also in Madrid and Cologne, in Paris and Bilbao, or in Lisbon and Amsterdam.} After that, sunrise line sweeps the British islands. In the afternoon, the terminator line sweeps Europe from Northeast to Southwest; the sun leaves Europe through the Southwestern tip of the Iberian Peninsula).

The preceding paragraph is the description of a natural phenomenon which is immutable. Contrastingly, societies must decide which time zone will clocks display and this decision changes winter sunrise and sunset local times, and then alter Fig.~\ref{fig:labor} (top). 

\begin{figure}
  \centering
    \includegraphics[width=\linewidth]{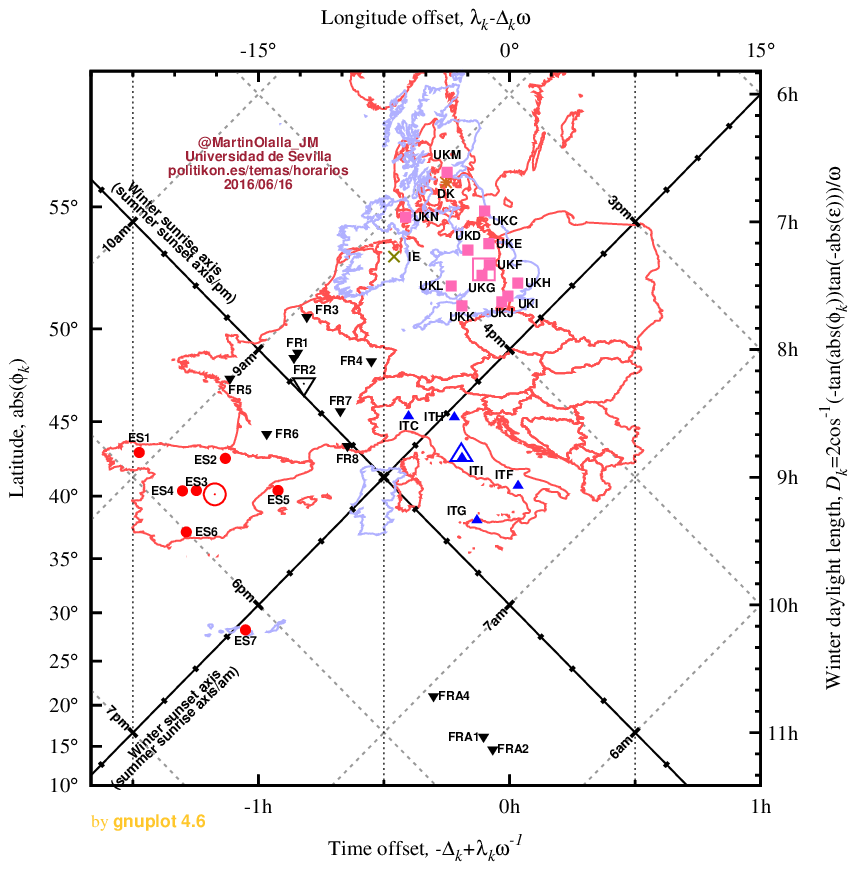}
  \caption{A collapsed version of Fig.~\ref{fig:mapaeuropa} where time zone is taken into account and the horizontal axis display time offset instead of longitude. The main consequence of that is the shifting of WET regions (Portugal, United Kingdom, Ireland and Canary Islands relative to CET regions. Larger open symbols display the weight-averaged latitude and offset for a country. Vertical axis displays time distance from local civil noon to local solar noon. Slanted grid display winter sunrise axis and winter sunset axis. The terminator at winter sunrise or sunset is perpendicular to its respective axes.}
  \label{fig:colapsa}
\end{figure}

Fig.~\ref{fig:colapsa} shows in the horizontal axis the time offset $\tau_i$ (see Eq.~(\ref{eq:2})) ---the time distance from local solar noon to local civil noon--- thus takes into account the time zone of regions. Hence the map shown in Fig.~\ref{fig:colapsa} is slightly different from a standard map, since $\Delta_i$ does not get the same value everywhere. Notice England comes over East Germany, Scotland mostly over Denmark and Ireland is over Netherlands.

Horizontal axis essentially display noon properties, the slanted grid shown on the plot display the solstitial sunrises and sunsets, which are opposite (winter sunrise matches to summer sunset).\footnote{Notice we are defining sunrises and sunsets by $z=\unit{0}{\degree}$, which is a simplification, and we are not taking into account daylight saving time.} Fig.~\ref{fig:colapsa} then shows the unique consequences of any time zone choice: it sets noon and it sets the extremal sunrises and sunsets.

Fig.~\ref{fig:colapsa} shows that with the present choice of time zones in Europe, winter sunrise progresses in Western Europe from \uhora{08}am to \uhora{09}am (local time), thus making eight o'clock (local time) a popular choice for getting to work. 

\begin{figure*}
  \centering
    \includegraphics[width=\textwidth]{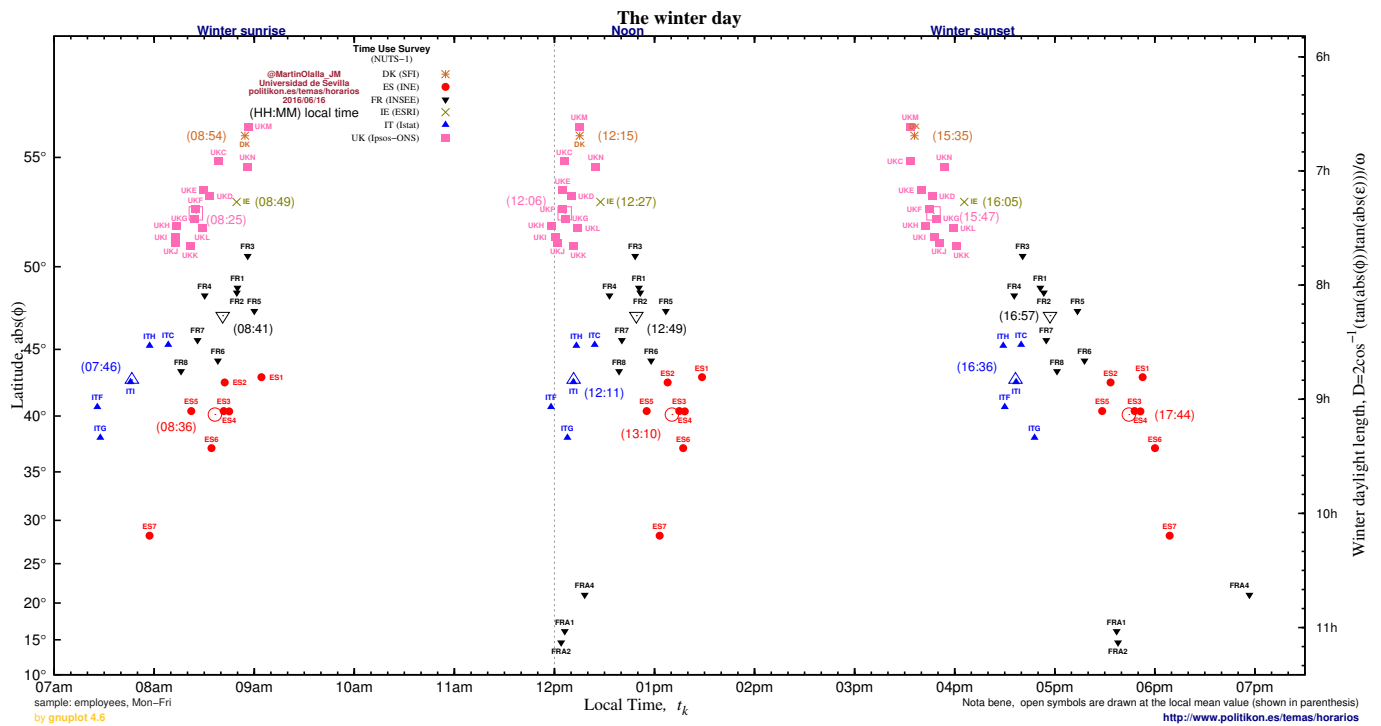}
  \caption{Winter sunrise, noon, and winter sunset local times plotted along with latitude as vertical coordinate. The similarities with Fig.~\ref{fig:labor} (top) are evident.}
  \label{fig:amanecer}
\end{figure*}

On the contrary, winter sunset lacks of a common local time in Western Europe. The terminator sweeps Europe from Northeast to Southwest and darkness steadily covers Denmark and United Kingdom first, then Ireland, then France, Italy and, finally, Spain. Fig.~\ref{fig:amanecer} mimics the top panel of Fig.~\ref{fig:labor} but shows local times for winter sunrise, noon, and winter sunset instead. The correspondence between these two figures are evident.

The interesting point here is that had France and Spain turned back to Western Europe Time zone at the end of World War Two, they would be placed over Italy, Austria and the Czech Republic in Fig.~\ref{fig:colapsa}. Hence, the latest sunrise time would happen one hour earlier and seven o'clock may have been a popular choice for the labor start time in Spain and, probably, in France. That is the choice in the United States of America ---which share latitude with the Iberian Peninsula\footnote{In Fig.~\ref{fig:colapsa} the East Coast of the United States would lie east of Italy, and winter sunrise occur at seven o'clock local time}. The reader should be aware of the trick: 07WET is exactly the same as 08CET.

The bottom line here is that by setting clocks to CET zone, France and Spain have prompted a common winter sunrise in Western Europe, Germany and the British Islands. Ultimately that choice leads to common labor start times (as observed in Fig.~\ref{fig:labor} (top)). As a consequence they have fuelled different winter sunset and labor end times (see also Fig.~\ref{fig:labor} (top) and Fig.~\ref{fig:amanecer}).\footnote{As a example, winter sunset is almost coincidental in Paris and London. Yet it is \uhora{03:45}pm local time in London, and \uhora{04:45}pm local time in Paris.} However these differences, though noticeable, are unimportant. On the contrary they are convenient as apparently, they only highlights the way human activity fits to solar activity (see Fig.~\ref{fig:labor} (bottom)). To put it shortly they tell us about a natural phenomenon.

Finally it should be stressed that only at temperate latitudes can society fit the human activity to the solar activity this way. To be specific, only if the winter day length is close to the average daily labor time ---which roughly speaking ranges in seven to eight hours per day and employed person---could society fit that way. If day length is too short ---at high enough latitude--- employed people have the necessity of starting and/or ending their duties before sunrise or after sunset. On the contrary, if day length is too long ---at low enough latitude--- there is no need to fit labor start and end times to winter sunrises or sunsets. 

\subsection{Noon, lunch break and day length}
\label{sec:lunch-break}

The preceding discussion focused on labor start and end times and the way they fit to the winter sunrise/sunset lines. Noon also deserves an explanation.

The reader should notice that labor activity fits to winter sunrise in the morning and to winter sunset in the afternoon. And they have opposite trends. There must be a cross-over somewhere in the day: it is at noon.

Labor noon time ---the time when half the total labor activity of a region has been already consumed--- correlates well with the winter sunrise, see Fig.~\ref{fig:labor} (bottom). Labor noon time comes anywhere from four to five hours after winter sunrise. This is likely telling us that morning labor is relatively similar across societies.\footnote{Albeit some minor differences which are evident in Fig.~\ref{fig:labor}, for instance Italy is delayed with respect to France in the morning, but advanced at noon.}

By comparison afternoon labor must be different. This comes out by simply considering that at labor noon times follow the sunrise trend ---the higher the latitude, the later the time--- while labor end times follow the opposite trend. The only way this is achieved by slowing down labor activity at low latitude regions. 

One way of so doing is having longer lunch break the longer your winter day length is. The rationale here is that winter day is long enough so as to exceed the average time spent at work. Hence longer intraday breaks are possible without resuming labor after any sunset of the year.

Another way of so doing is just by burning work in the morning, so that employed persons need not to come back to work in the afternoon. This scenario is only possible at low latitude since as latitude decreases, winter day length increases and the time distance from sunrise to noon increases as well. Longer mornings can accommodate larger amounts of work. 

In fact, data in Fig.~\ref{fig:labor} and Fig.~\ref{fig:comer} shows that only in Italy and Spain (the southern most countries of the analysis) does the labor noon time go well before lunch time. Contrastingly in Denmark, Ireland and United Kingdom labor noon time and lunch time differ in few minutes, while in France lunch time comes well before labor noon time.

\subsection{Hours ticked by}
\label{sec:way-clocks-tic}

Clocks and longitude are linked to each other since the latter can only be measured with the help of the former. 

Clocks were then linked to noon in modern times. Noon, very conveniently, has a regular period of twenty four hours year round. But nobody is perfect. As we set our clocks to noon ---the most regular phenomenon--- we lost track of sunrises and sunsets ---the most sensible phenomena--- which primarily set human activity.

Hence, the vast myriad of labor end times observed in the top panel of  Fig.~\ref{fig:labor} (top) can easily hide the natural phenomenon observed in the bottom panel.

This is also speaking about a fundamental issue when dealing with clocks. Although meridians are very convenient, it should not be forgotten that points along a meridian are different to each other in a fundamental time property: winter day length.

As a running example we take the cities of London and Castellon (in Eastern Spain), which share the prime meridian. Yet, noon at Castellon comes at least 4h45m after sunrise; while at London noon can come as low as 3h45m after sunrise, a phenomenon impossible to be observed in the Spanish city. That makes a difference when social time schedules are build up. Castellon and London do share daily noon; but they do not share winter sunrise and hence, they will not share labor start time. 

On the contrary, points sharing a latitude are all identical in some way. For sure they differ in the longitude. But longitude is just a matter of time: sooner or later whatever happened somewhere, will happen somewhere else along a parallel. For instance, whatever happened in Naples, will happen one hour later in Castellon.

Physicists, geographers, astronomers are fond of meridians, transits and noons which remain highly unnoticeable for the average person. People are quite more sensible to sunrise and sunset. 

\section{Conclusion}
\label{sec:conclusion}

Winter day provides an explanation for the way some European societies develop their daily activities. The rationale is if you survive to the worst case scenario of the year ---the shortest day of the year--- you will manage to survive to any other day. That would be true at temperate latitudes where winter day-length is similar to the daily average working time.

In the morning, differences are rare because societies come from a common fundamental rest state, and the only activities an employed person does are related to working. 

In the afternoon, employed persons develop activities non related to work. The winter sunset apparently triggers a myriad of social processes which include the end of labor ---somehow near to the winter sunset---, coming back home ---which happens some two to three hours after winter sunset--- or having dinner ---three to four hours after winter sunset---.

Since distance to winter sunrise or winter sunset can not be easily measured by clocks, this facts mostly remain unnoticed. People usually do not care why they get to work at eight o'clock in the morning, or why they have dinner at eight o'clock in the morning. 

From a scientific point of view the point here to emphasize is that times as provided by clock ---say, local times $t_i$--- must be compared to each other once time zone $\Delta_i$, longitude $\lambda_i$ and latitude $\phi_i$ are taken into account. These parameters enter ---together with Earth's angular speed of rotation $\omega$ and obliquity $\varepsilon$--- in the equation that provides time distance to winter sunrise (positive branch) and winter sunset (negative branch) $\delta_i$:
\begin{equation}
  \label{eq:3}
  \delta_i=t_i-\Delta_i-\unit{12}{\hour}+\frac{1}{\omega}\left(\lambda\pm\cos^{-1}(\tan(|\phi_k|)\tan(|\varepsilon|))\right)
\end{equation}
\section{Acknowledgments}
The author acknowledges the institutions of Denmark, Spain, France, Ireland, Italy and United Kingdom that provided the surveys and, thanks them for releasing the microdata to the scientific community.

The author also acknowledges to Politikon (\url{http://www.politikon.es/}) ---a think tank based in Spain--- for hosting most of the results and discussions related to this work. The author also wants to thank to his editor, Octavio Medina.

The author thanks Dr. Jos\'e Fern\'andez-Albertos at Centro de Ciencias Humanas y Sociales from the Spanish Centro Superior de Investigaciones Cient\'\i ficas for some fruitful discussions through Twitter social network.

\appendix

\section{At home}
\label{sec:at-home}

Time Use Surveys not only track daily activities but also track locations. As simple analysis of the location where activities occur allows us to  track whether the employee is at home or not and an ``activity'' which computes the shares of employed persons at home. From this statistics it is possible to compute times ascribed to leaving home or coming home.

It should be mentioned that leaving home in the morning is mostly related to going to work as soon as possible, while coming back home in the afternoon is not necessarily. People often develop leisure activities ---or more generally non-working activities--- only in the afternoon.

\begin{figure*}
  \centering
    \includegraphics[width=\textwidth]{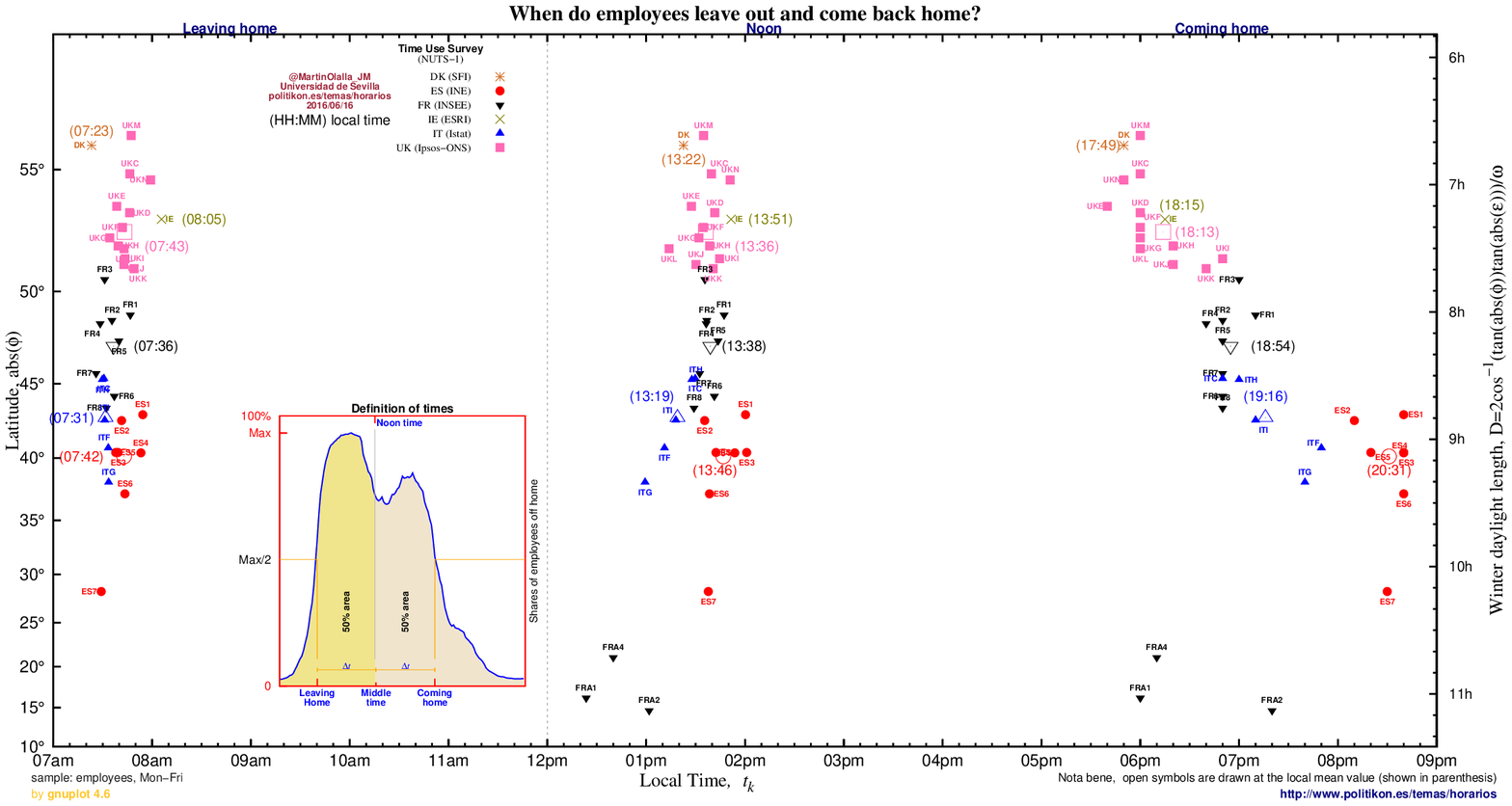}

  \includegraphics[width=\textwidth]{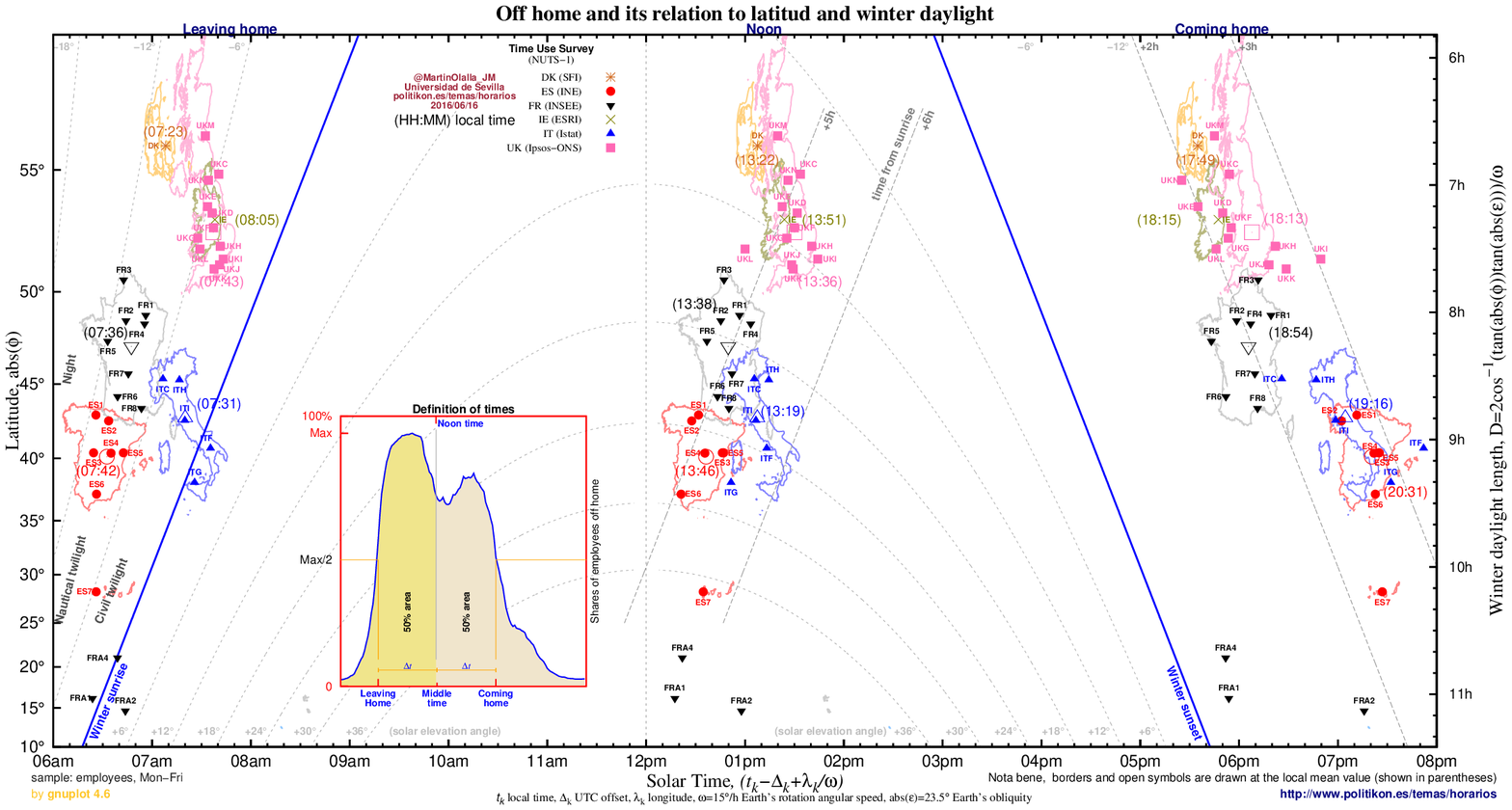}

  \caption{Top, local times for leaving home, coming home and the noon condition. As previously observed morning time are rather similar in Europe while the afternoon times are variable. Bottom the same plot but with the solar time in the $x$-axis. Data are comprehensively described with the coming home condition located from two to three hours after winter sunset.}
  \label{fig:casa}
\end{figure*}

Fig.~\ref{fig:casa} shows the local (top) and solar (bottom) times corresponding to this ``activity''. They show similar trends to those observed in Fig.~\ref{fig:labor}.

As observed in the bottom panel, employed persons are back at home some two-three hours after winter sunset.

\section{Eating times}
\label{sec:eating-times}

\begin{figure*}
  \centering
    \includegraphics[width=\textwidth]{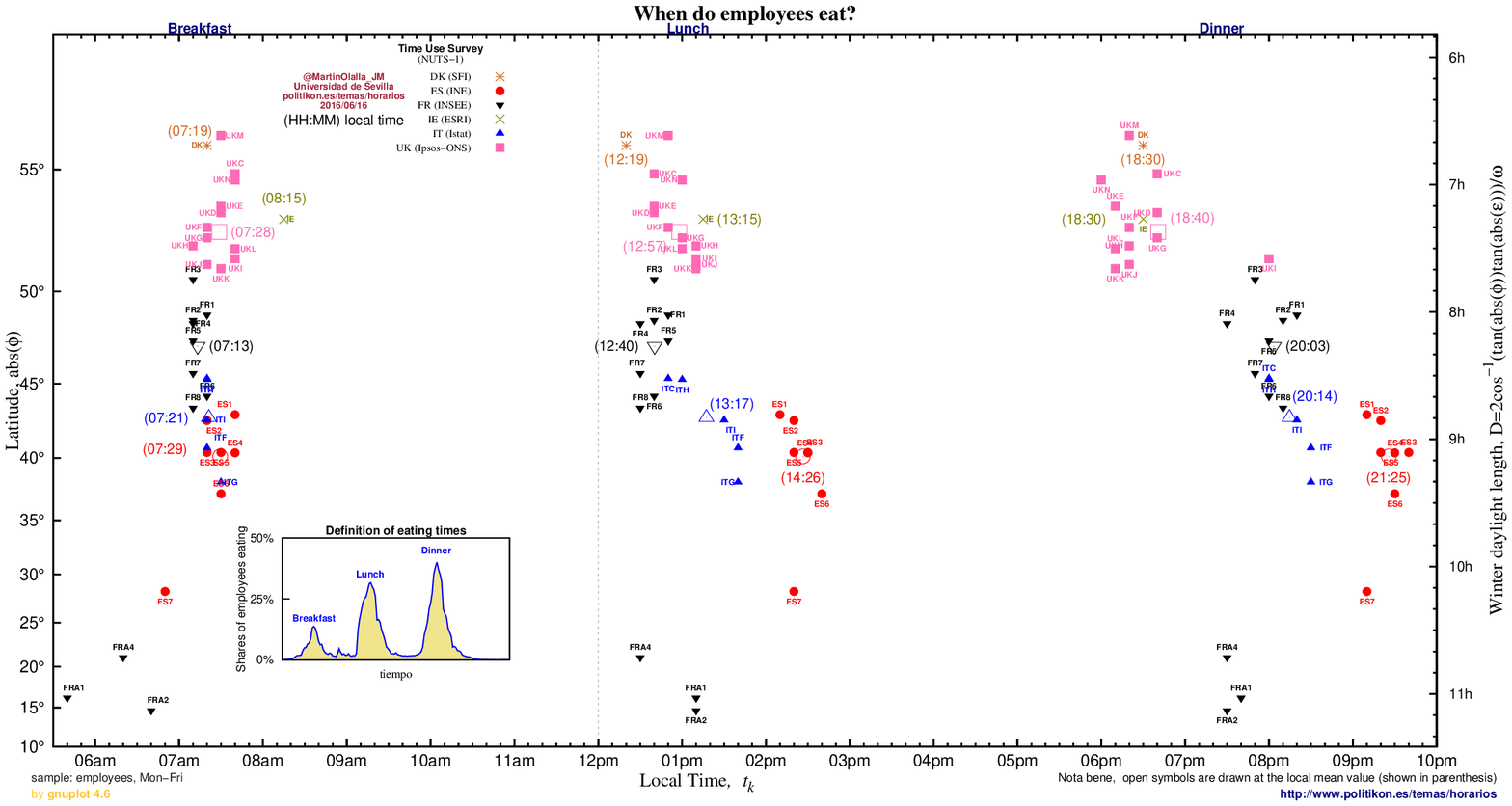}

  \includegraphics[width=\textwidth]{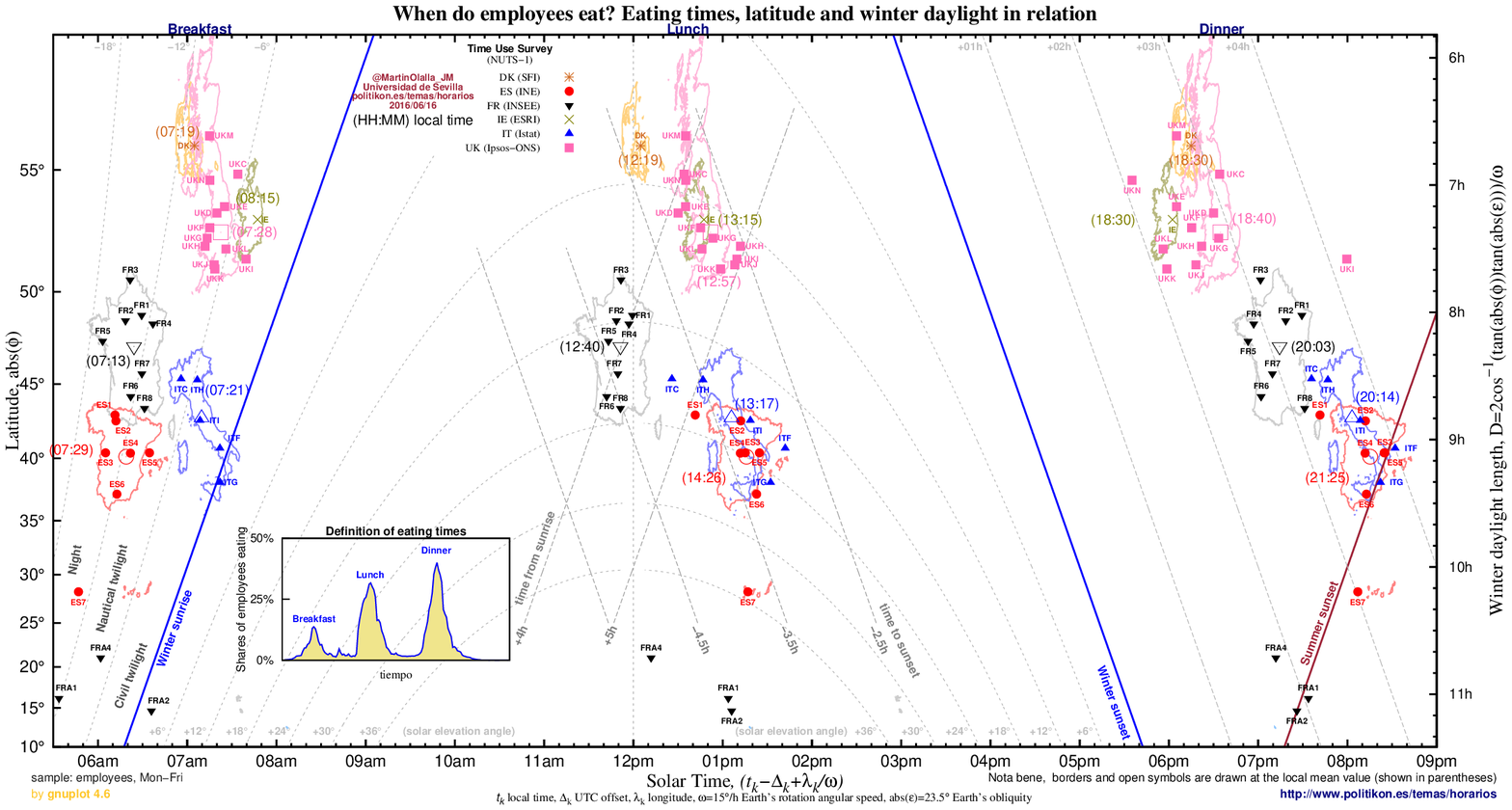}

  \caption{Top, local times for the breakfast, lunch and dinner determined by the positions of the bursts observed in the inset. As in Fig.~\ref{fig:labor} breakfast occurs in a narrow range of times, while lunch and dinner times progressively get wider ranges. Bottom, solar times for breakfast, lunch and dinner and their relation to the winter day. Summer sunset line is delayed by one hour due to daylight saving time.}
  \label{fig:comer}
\end{figure*}

Fig.~\ref{fig:comer} shows the eating local (top) and solar (bottom) times. Local times display the usual trend observed in Fig.~\ref{fig:labor} and explained in Sec.~\ref{sec:results}.

Solar times comprehensively describe data. Lunch time occur in Europe three hours before winter sunset, except in France, and dinner time mostly in the range three to four hours after winter sunset. An interesting question to answer is what does lunch mean? Two answers seem possible: (1) eating at noon, (2) eating well before (winter) sunset.

A possible explanation for the French anomaly deals with the time zone change enforced in 1940 amidst World War Two. The point here is that lunch times could have not changed since 1940 and thus they are located off the main trend. On the contrary, in Spain ---which carries the same history of time zone changes--- lunch times likely got delayed after clocks were set one hour ahead in 1940, thus offsetting the time zone change.

This idea can be tested by removing the time zone change in France and Spain (say, making their time zone equal to Western European Time $\Delta_k=0$ when computing $t^{\star}_k$). In so doing French and Spanish data will move one hour rightward. French data would then lie on the main trend thus suggesting that their lunch times are those prior to World War Two. Contrastingly, Spanish data would lie way off the main trend and anomalously close to the winter sunset.

Bottom panel of Fig.~\ref{fig:comer} also display the summer sunset ---shifted by one hour due to daylight saving time--- which is almost coincidental with the dinner time at Southeastern Italy and Spain. Hence with little effort, dinner times in Spain and Italy can be delayed so that dinner may become ``eating at night''. Of course it is not possible to achieve this task in France, United Kingdom, Ireland, Denmark or Northwestern Spain and Italy where having dinner before sunset is usual.

\section{Sleep times}
\label{sec:wakeful-times}

\begin{figure*}
  \centering
    \includegraphics[width=\textwidth]{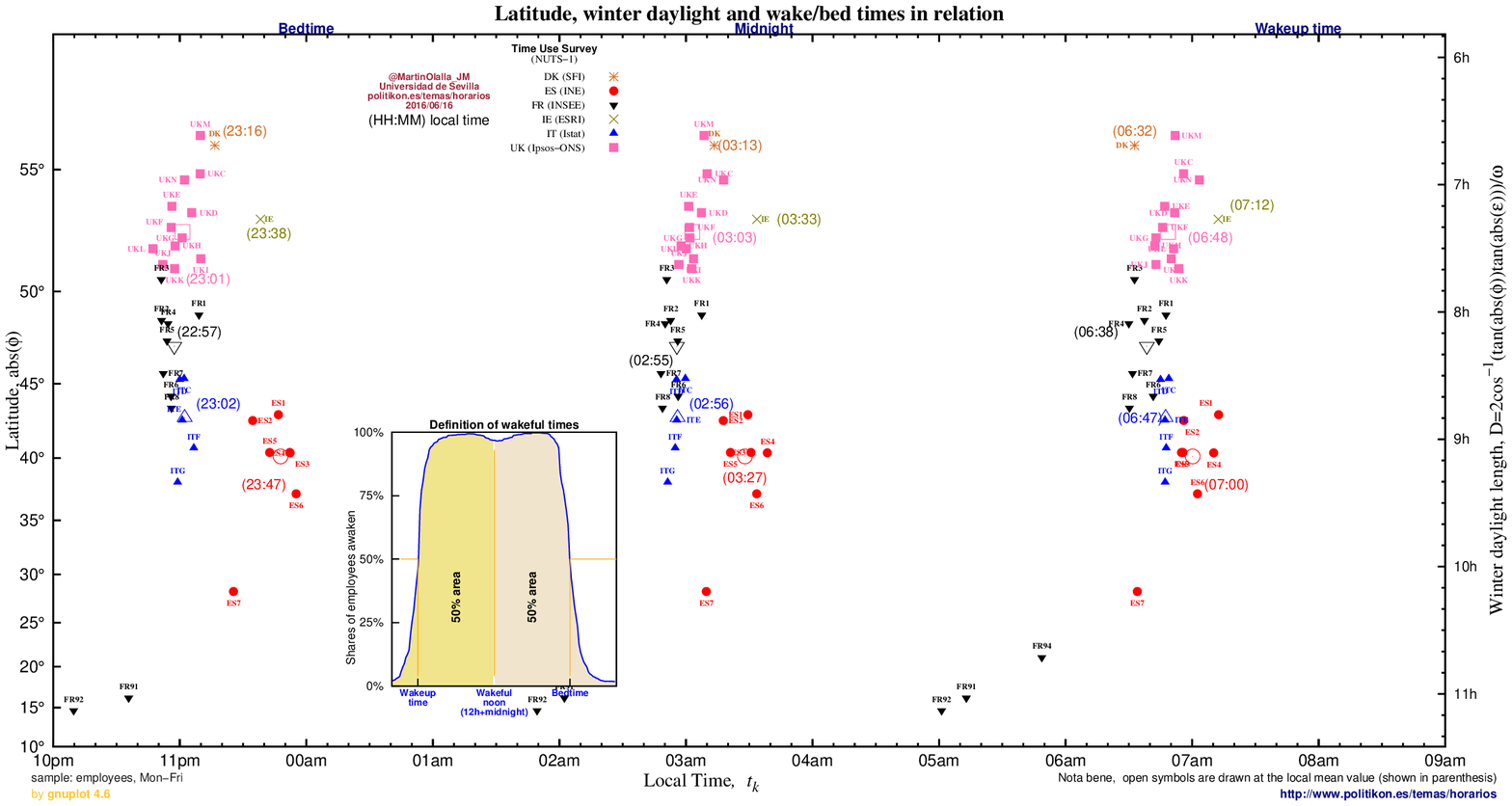}

  \includegraphics[width=\textwidth]{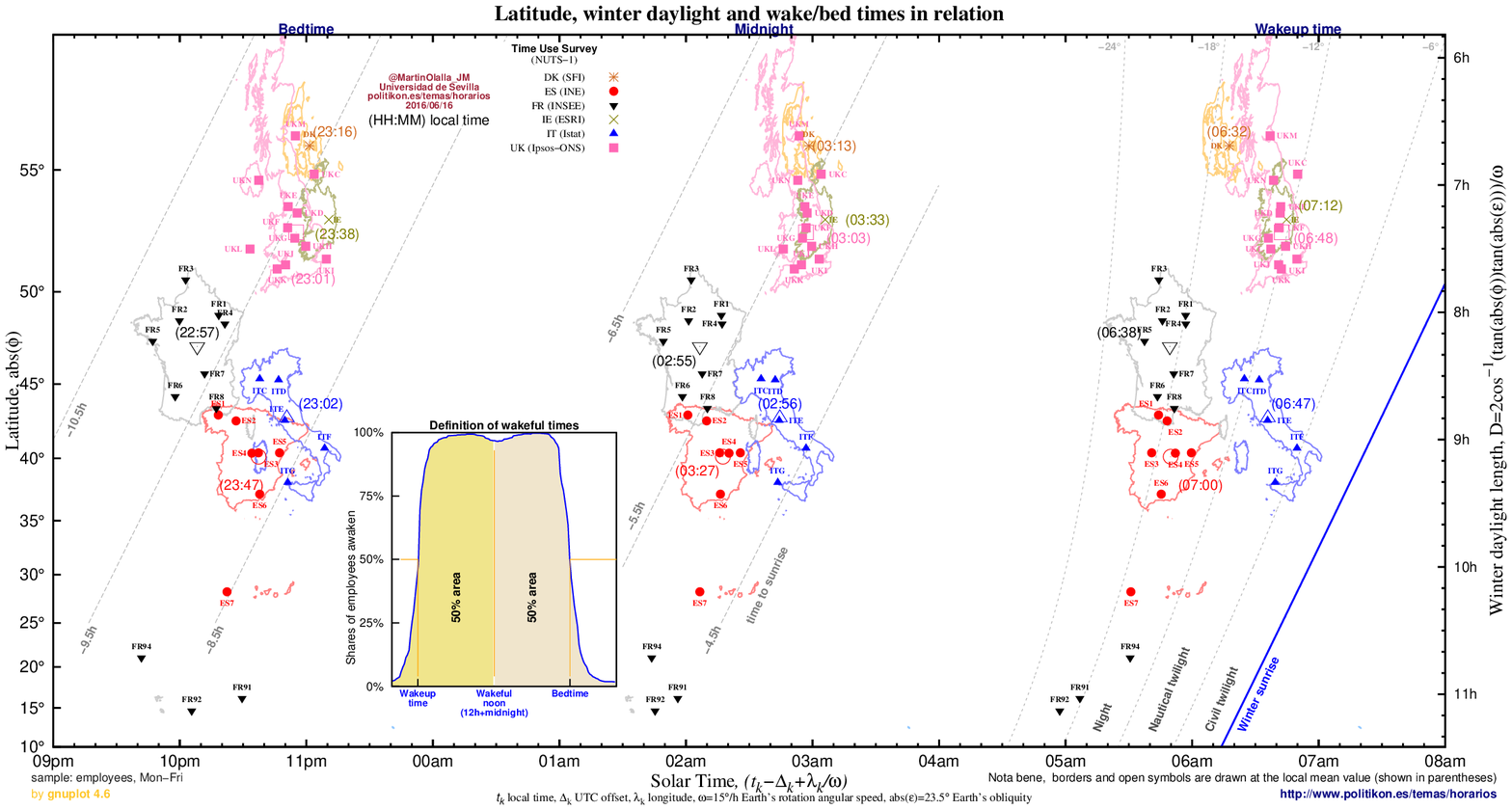}

  \caption{Top, local times for the sleeping activity (bedtimes, noon and wake-up time) and latitude. Bottom, the corresponding solar times. Sleep times do not follow the trends observed elsewhere in this work. They play the role of a crossover from the positively slanted morning times to the negatively slanted afternoon times.}
  \label{fig:dormir}
\end{figure*}

Times related to the sleeping activity can be observed in Fig.~\ref{fig:dormir} where the top panel displays local times and the bottom panel displays solar times.

The interesting point here is that sleep times can hardly be related to winter sunrise or winter sunset, albeit the trivial idea that wake-up time are close to winter sunrise. 

The issue here is that trends must be reversed again. So bedtimes play the same role of crossing-over as lunch-break was playing at noon. For instance northern countries exhibit earlier dinner solar times as compared to southern countries (see Fig.~\ref{fig:comer} (bottom)) but their bed times are not that early.

Nonetheless sleeping times are similar from survey to survey and range from \uhora{7h25m} per day and employed person in Denmark to \uhora{7h45m} per day and employed person in United Kingdom.


\end{document}